\documentstyle[12pt]{article}
\def\PR #1 #2 #3 {Phys.~Rev.~{\bf #1}, #2 (#3)}
\def\PRL #1 #2 #3 {Phys.~Rev.~Lett.~{\bf #1}, #2 (#3)}
\def\PRD #1 #2 #3 {Phys.~Rev.~D~{\bf #1}, #2 (#3)}
\def\PLB #1 #2 #3 {Phys.~Lett.~{\bf B#1}, #2 (#3)}
\def\NPB #1 #2 #3 {Nucl.~Phys.~{\bf B#1}, #2 (#3)}
\def\RMP #1 #2 #3 {Rev.~Mod.~Phys.~{\bf #1}, #2 (#3)}
\def\ZPC #1 #2 #3 {Z.~Phys.~C~{\bf #1}, #2 (#3)}

\newcommand{\pslash}{p\!\!\!/} 
\newcommand{\lqcd}{\Lambda_{\rm QCD}}

\begin{document}
\include{psfig}
\begin{titlepage}

\rightline{hep-ph/9612329} 
%\rightline{ILL-(TH)-96-9}
\medskip
\rightline{December 1996}
\bigskip\bigskip

\begin{center} {\Large \bf Top-quark pole mass} \\
\bigskip\bigskip\bigskip\bigskip
{\large\bf Martin C.~Smith and Scott S.~Willenbrock} \\ 
\medskip 
Department of Physics \\
University of Illinois \\ 1110 West Green Street \\  Urbana, IL\ \ 61801 \\
\bigskip 
\end{center} 
\bigskip\bigskip\bigskip

\begin{abstract}
The top quark decays more quickly than the strong-interaction time
scale, $\lqcd^{-1}$, and might be expected to escape the effects of
nonperturbative QCD.  Nevertheless, the top-quark pole mass, like the
mass of a stable heavy quark, is ambiguous by an amount proportional
to $\lqcd$.
\end{abstract}

\end{titlepage}

\newpage

\section{Introduction}

\indent\indent The mass of the recently-discovered top quark
\cite{TOP} has been measured with impressive accuracy, $m_t=175\pm 6$
GeV \cite{MASS}, by the CDF and D0 experiments at the Fermilab
Tevatron.  The uncertainty will be reduced even further, to perhaps
1-2 GeV, with additional running at the Tevatron \cite{TEV2000}, or at
the CERN Large Hadron Collider \cite{ATLAS}.  High-energy $e^+e^-$
\cite{NLC} or $\mu^+\mu^-$ \cite{MUON} colliders operating at the
$t\bar t$ threshold hold the promise of yet more precise measurements
of $m_t$, to 200 MeV or even better.

With such increasingly-precise measurements on the horizon, it is
important to have a firm grasp of exactly what is meant by the
top-quark mass.  Thus far the top-quark mass has been experimentally
defined by the position of the peak in the invariant-mass distribution
of the top-quark's decay products, a $W$ boson and a $b$-quark jet
\cite{MASS}.  This closely corresponds to the pole mass of the top
quark, defined as the real part of the pole in the top-quark
propagator.  The propagator of a top quark with four-momentum $p$ has
a pole at the complex position $\sqrt {p^2} = m_{pole} -
\frac{i}{2}\Gamma$, and yields a peak in the $Wb$ invariant-mass
distribution (for experimentally-accessible real values of $p$) when
$\sqrt {p^2} \approx m_{pole}$.

The pole mass of a stable quark is well-defined in the context of
finite-order perturbation theory \cite{T}.  However, the all-orders
resummation of a certain class of diagrams, associated with ``infrared
renormalons'', indicates that the pole mass of a stable
heavy\footnote{Heavy here means $m \gg \lqcd$.} quark is ambiguous by
an amount proportional to $\lqcd$, as a result of nonperturbative QCD
\cite{BSUV,BB}.  Physically, this is a satisfying result, because we
believe that quarks are permanently confined within hadrons,
precluding the unambiguous definition of a quark pole mass \cite{BU}.

The top quark decays very quickly, having a width $\Gamma \approx
{1.5\;{\rm GeV}}$, approximately an order of magnitude greater than
the strong-interaction energy scale $\lqcd \approx 200\;{\rm MeV}$.
Such a short lifetime means that the top quark decays before it has
time to hadronize \cite{K,BDKKZ,OR}.  The large top-quark width can
act as an infrared cutoff, potentially insulating the top quark from
the effects of nonperturbative QCD \cite{FK,SP,KS}.

Given this information, one might expect the top-quark pole mass to be
free of the ambiguities associated with nonperturbative QCD.  The
purpose of this article is to demonstrate that this is not the case.
The top-quark pole mass, like the mass of a stable heavy quark, is
unavoidably ambiguous by an amount proportional to $\lqcd$.  We
demonstrate this in two ways, first by a general argument using
$S$-matrix theory, second by a consideration of infrared renormalons.
The ambiguity in the pole mass in the specific context of the $Wb$
invariant-mass distribution is discussed at the end of the next
section.

\section{General Argument}

\indent\indent Consider a scattering process with asymptotic states
consisting of stable particles.  We first ask if it is possible for
the scattering amplitude to have a pole at the mass of a stable quark.
This would correspond to a quark propagator connecting two
subamplitudes, as depicted in Fig.~1; the pole in the quark propagator
would correspond to the pole in the amplitude.  Such a configuration
is impossible, however: the subamplitudes which the quark propagator
connects have external states which are color singlets (due to
confinement), while the quark is a color triplet, so color is not
conserved.  Thus there cannot be a pole in the amplitude at the quark
mass.

\begin{figure}
\centerline{\psfig{figure=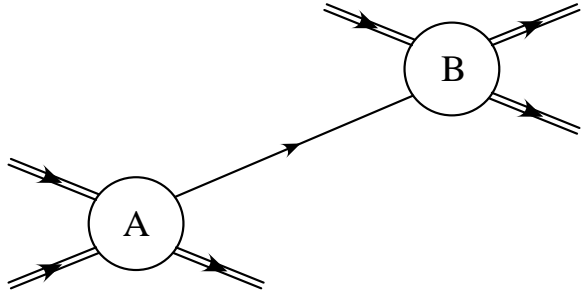,width=3in}}
\caption{\footnotesize A scattering amplitude factorizes when an
internal propagator is near its pole.  The external lines represent
color-singlet asymptotic states.}
\end{figure}

This argument applies equally well to an unstable quark, such as the
top quark. The fact that the quark is unstable evidently plays no role
in the argument; it only shifts the imagined pole in the propagator
into the complex plane. As in the case of a stable quark, there cannot
be a pole in the amplitude, regardless of how short-lived the quark.
In particular, the fact that the top-quark lifetime is much less than
$\lqcd^{-1}$ is irrelevant.

There is another way to understand why the short top-quark lifetime is
irrelevant.  Let us return to Fig.~1, and again consider first the
case of a stable quark.  Imagine that there is a pole in the amplitude
at the quark mass.  Near the pole, the scattering amplitude would
factorize into the production of the stable quark by scattering
subprocess A, followed by its propagation over a large proper time,
and concluding with its participation in scattering subprocess B.
Thus the quark could be considered as an asymptotic state.  This
demonstrates that the poles in the scattering amplitude of a theory
correspond to its asymptotic states \cite{ELOP1}.  Since quarks are
not asymptotic states, due to confinement, there cannot be a pole at
the quark mass.

A similar argument applies to an unstable quark, such as the top
quark.  The imagined pole position is now located at a complex value.
Because the scattering amplitude is an analytic function, the analytic
continuation to complex momentum is well-defined.  Near the pole, the
scattering amplitude would factorize as before, although this would no
longer correspond to a true physical process since the top-quark would
have complex momentum \cite{ELOP2}.  The top quark would propagate
over a large proper time, and could not escape confinement.  There
would be no asymptotic top-quark state, and hence no pole.

We are left with the following physical picture.  A state with
momentum near its pole corresponds to a long-lived particle,
regardless of whether the pole is real or complex.  If the particle is
colored, it will be confined, preventing an unambiguous definition of
the pole mass of the particle.\footnote{This picture also implies that
there are poles associated with hadrons containing a top quark, but
these poles are far from the real axis, due to the large top-quark
width.}

These arguments imply that the nonperturbative aspect of the strong
interaction will stand in the way of any attempt to unambiguously
extract the top-quark pole mass from experiment.  For example,
consider the extraction of the pole mass from the peak in the $Wb$
invariant-mass distribution.  In perturbation theory, the final state
is a $W$ and a $b$ quark, as depicted in Fig.~2(a).  However, the $b$
quark manifests itself experimentally as a jet of colorless hadrons,
due to confinement.  At least one of the quarks which resides in these
hadrons comes from elsewhere in the diagram, and cannot be considered
as a decay product of the top quark, as depicted in Fig.~2(b).  This
leads to an irreducible uncertainty in the $Wb$ invariant mass of
${\sl O}(\lqcd)$, and hence an ambiguity of this amount in the
extracted top-quark pole mass.

\begin{figure}
\centerline{\psfig{figure=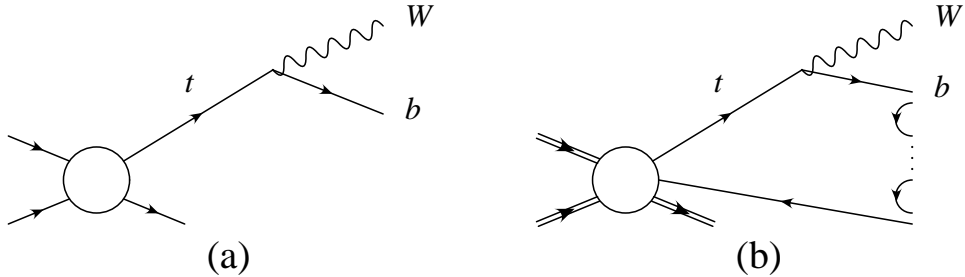,width=5in}}
\caption{\footnotesize The production and decay of a top quark in (a)
perturbation theory, and (b) nonperturbatively.}
\end{figure}

\section{Infrared Renormalons}
 
\indent\indent We now turn to an investigation of the top-quark pole
mass from the perspective of infrared renormalons.  We first review
the argument which demonstrates the existence of a renormalon
ambiguity in the pole mass of a stable heavy quark \cite{BSUV,BB}.  We
then extend the argument to take into account the finite width of the
top quark.  Finally, we investigate the existence of a renormalon
ambiguity in the top-quark width itself.

The pole mass of a quark is defined by the position of the pole in the
quark propagator.  The propagator of a quark of four-momentum $p$ is
\begin{equation}
D(\pslash) = \frac{i}{\pslash - m_R - \Sigma(\pslash)}
\end{equation}
where $m_R$ is a renormalized short-distance mass,\footnote{By
short-distance mass we mean a running mass (such as the $\overline{\rm
MS}$ mass) evaluated at a scale much greater than $\lqcd$.}  and
$\Sigma(\pslash)$ is the renormalized one-particle-irreducible quark
self-energy.  The equation for the position of the pole is
\begin{equation}
\pslash_{pole} = m_R + \Sigma(\pslash_{pole}) \; .
\label{pole}
\end{equation} 
This is an implicit equation for $\pslash_{pole}$ that can be solved
perturbatively.  We first work to leading order in $\alpha_s$, which
gives
\begin{equation}
\pslash_{pole} = m_R + \Sigma^{(1)}(m_R)
\label{pole1}
\end{equation}
where $\Sigma^{(1)}(m_R)$ is the one-loop quark self-energy shown
Fig.~3(a).  This quantity is real, so the pole position
is real.

\begin{figure}
\centerline{\psfig{figure=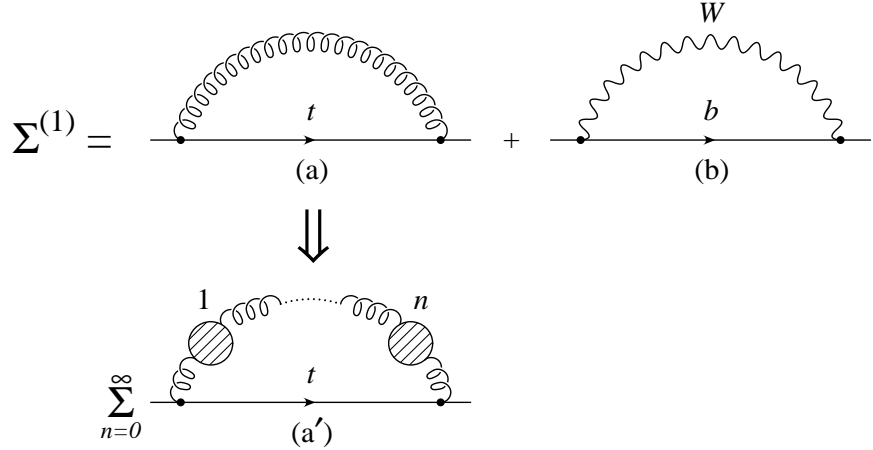,width=5in}}
\caption{\footnotesize Diagrams contributing to the top-quark
self-energy at leading order in $\alpha_s$ and $\alpha_W$.  Fig.~(a$^\prime$)
replaces Fig.~(a) when summing to all orders in $\beta_0\alpha_s$.}
\end{figure}

Renormalons arise from the class of diagrams generated by the
insertion of $n$ vacuum-polarization subdiagrams into the gluon
propagator in the one-loop self-energy diagram, as shown in
Fig.~3(a$^\prime$). One can express this as
\begin{equation}
\Sigma^{(1)}(m_R,a) = \frac{16m_R}{3\beta_0} \sum_{n=0}^{\infty} c_n
a^{n+1}
\label{sigma}
\end{equation}
where 
\begin{equation}
a \equiv \frac{\beta_0\alpha_s(m_R)}{4\pi}
\end{equation}
and $\beta_0$ is the one-loop QCD beta-function coefficient, $\beta_0
\equiv 11-(2/3)N_f$.  Formally, these are the dominant QCD corrections
in the ``large-$\beta_0$'' limit.  Thus $\Sigma^{(1)}(m_R,a)$ in
Eq.~(\ref{sigma}) is calculated at leading order in $\alpha_s$, but to
all orders in $a$.

For large $n$ the coefficients $c_n$ grow factorially, and are given
by \cite{NNA,BBB,PS}
\begin{equation}
c_n \stackrel{n\to\infty}{\rightarrow} e^{-C/2} 2^n n!
\label{coeff}
\end{equation}
where $C$ is a finite renormalization-scheme-dependent
constant.\footnote{In the $\overline{\rm MS}$ scheme, $C=-5/3$.}  The
series in Eq.~(\ref{sigma}) is therefore divergent.  One can attempt
to sum the series using the technique of Borel resummation \cite{tH}.
The Borel transform (with respect to $a$) of the self-energy is
obtained from the series coefficients, Eq.~(\ref{coeff}), via
\begin{equation}
\widetilde\Sigma^{(1)}(m_R,u) = \frac{16}{3\beta_0} m_R
\sum_{n=0}^{\infty} \frac{c_n}{n!}\,u^n
\end{equation}
where $u$ is the Borel parameter.  Because the coefficients $c_n$ are
divided by $n!$ in the above expression, the series has a finite
radius of convergence in $u$, and can be analytically continued into
the entire $u$ plane.  The self-energy is then reconstructed via the
inverse Borel transform, given formally by
\begin{equation}
\Sigma^{(1)}(m_R,a) = \int_0^\infty du\, e^{-u/a}
\widetilde\Sigma^{(1)}(m_R,u)
\label{inverse}
\end{equation}

The integral in Eq.~(\ref{inverse}) is only formal, because the Borel
transform of the quark self-energy possesses singularities on the real
$u$-axis, which impede the evaluation of the integral.  These
singularities are referred to as infrared renormalons because they
arise from the region of soft gluon momentum in Fig.~3(a$^\prime$).
The series for the self-energy in Eq.~(\ref{sigma}) is therefore not
Borel summable.

The divergence of the series for the self-energy is governed by the
infrared renormalon closest to the origin, which lies at $u=1/2$.
This renormalon is not associated with the condensate of a local
operator, so it cannot be absorbed into a nonperturbative redefinition
of the pole mass \cite{BSUV,BB}.  Instead, one can choose some {\it ad
hoc} prescription to circumvent the singularity in the integral.  The
difference between various prescriptions is a measure of the ambiguity
in the pole mass.  Estimating the ambiguity as half the difference
between deforming the integration contour above and below the
singularity gives \cite{BB}
\begin{equation}
\delta m_{pole} \sim \frac{8\pi}{3\beta_0} e^{-C/2} \lqcd
\end{equation}
so the pole mass is ambiguous by an amount proportional to $\lqcd$.

We now include the $O(\alpha_W)$ contribution to the top-quark
self-energy shown in Fig.~3(b).  The pole position is still given by
Eq.~(\ref{pole1}), but where $\Sigma^{(1)}(m_R)$ includes both
Figs.~3(a) and (b).  Since Fig.~3(b) has an imaginary part, the pole
moves off the real axis.  The imaginary part of the one-loop pole
position defines the tree-level top-quark width via \mbox{${\rm Im} \,
\Sigma^{(1)}(m_R) \equiv -\frac{1}{2} \Gamma_{tree}$.}  As before, to
extend the calculation to all orders in $a$, we replace Fig.~3(a) by
Fig.~3(a$^\prime$).  This contribution to the pole mass remains the
same as for a stable quark, and has the same renormalon ambiguity.  At
leading order in $\alpha_W$, the infrared renormalons do not know
about the top-quark width.

The $O(\alpha_s)$ contribution to the top-quark self-energy learns
about the top-quark width if one works to all orders in $\alpha_W$,
via a Schwinger-Dyson representation \cite{SD}, as shown in Fig.~4.
The circles on the internal propagators and the vertex in Figs.~4(a)
and (b) represent the weak corrections to all orders in
$\alpha_W$.\footnote{The circles in Fig.~4(b) also contain one power
of $\alpha_s$.}  We wish to solve for the pole position as given by
Eq.~(\ref{pole}).  We denote the pole position at zeroth order in
$\alpha_s$, but to all orders in $\alpha_W$, by the complex value $M$,
with ${\rm Im} \, M\, \equiv -\frac{1}{2}\Gamma$, where $\Gamma$ is
the top-quark width to all orders in $\alpha_W$.  At leading order in
$\alpha_s$, the pole position is then given by
\begin{equation}
\pslash_{pole} = m_R + \Sigma(M)
\end{equation}
where $\Sigma(M)$ is given by Figs.~4(a) and (b).  Again, we extend
this calculation to all orders in $a$ by making $n$
vacuum-polarization insertions in the gluon propagator, as depicted in
Fig~4(a$^\prime$).  This yields a series in $a$, which we denote by
$\Sigma(M,a)$ in analogy with Eq.~(\ref{sigma}).  To investigate
whether the width might cut off the infrared renormalons generated by
these diagrams, we need only consider the contribution of soft gluons.
In the limit of vanishing gluon momentum, the internal propagator
reduces to $Z/(\pslash - M)$, where $Z$ is the
wavefunction-renormalization factor.  The Ward identity tells us that,
in this same limit, the dressed vertex is simply $Z^{-1}$.  Thus, in
the infrared limit, $\Sigma(M,a)$ is formally identical to
$\Sigma^{(1)}(m_R,a)$ with $m_R$ replaced by $M$ everywhere.  The
infrared renormalons, which are associated with the Borel transform
with respect to $a$, are unaffected.  The width does not act as an
cutoff for infrared renormalons, despite the fact that it is much
greater than $\lqcd$.  We conclude that the pole mass of the top quark
is ambiguous by an amount proportional to $\lqcd$, just as for the
case of a stable quark.

\begin{figure}
\centerline{\psfig{figure=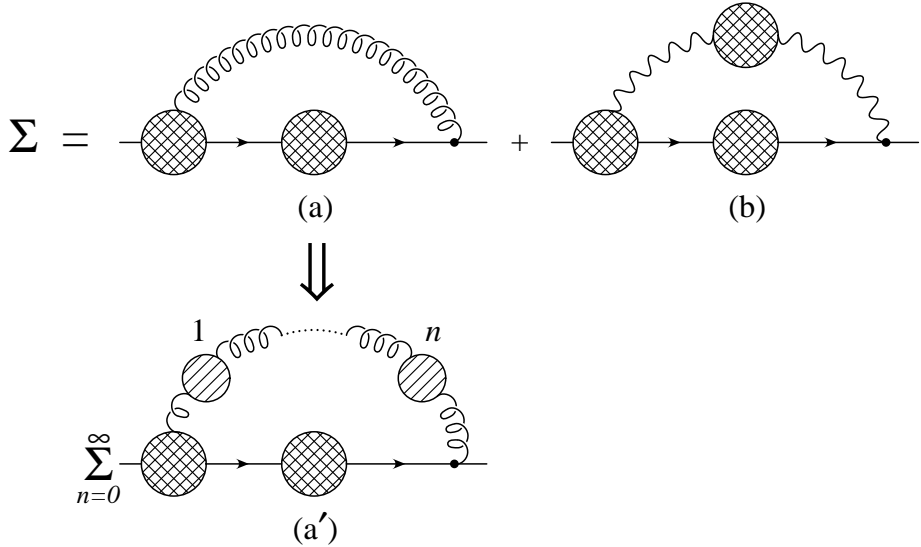,width=5in}}
\caption{\footnotesize Diagrams contributing to the top-quark self
energy at leading order in $\alpha_s$, but to all orders in
$\alpha_W$.  Fig.~(a$^\prime$) replaces Fig.~(a) when summing to all
orders in $\beta_0\alpha_s$.}
\end{figure}

We next ask whether the top-quark width suffers from a similar
renormalon ambiguity.  Because the first-order calculation yields the
top-quark width at tree level only, it is insufficient to address this
question.  The solution to Eq.~(\ref{pole}) at $O(\alpha_W\alpha_s)$
is
\begin{eqnarray}
\pslash_{pole} & = & m_R + \Sigma(m_R + \Sigma(m_R)) \nonumber \\ & =
& m_R + \Sigma^{(1)}(m_R) + \Sigma^{(2)}(m_R) +
\Sigma^{(1)\prime}(m_R)\Sigma^{(1)}(m_R)
\label{pole2}
\end{eqnarray}
where the superscripts on $\Sigma$ indicate the order at which it is
to be evaluated.  The imaginary part of this equation (times $-1/2$)
defines the top-quark width at $O(\alpha_W\alpha_s)$.

One may calculate the imaginary part of Eq.~(\ref{pole2}) using the
Cutkosky rules.  This reduces to the calculation of the QCD correction
to the process $t\to Wb$.\footnote{The term involving
$\Sigma^{(1)\prime}(m_R)$ corresponds to the wavefunction
renormalization of the top quark.}  The presence of renormalons in
this process was investigated in Refs.~\cite{NNA,BBB}.  If the width
is expressed in terms of the pole mass, then it has an infrared
renormalon at $u=1/2$, corresponding to an ambiguity proportional to
$\lqcd$.  However, if the width is expressed in terms of a
short-distance mass, such as the $\overline{\rm MS}$ mass, there is no
renormalon at $u=1/2$, and hence no ambiguity proportional to $\lqcd$.

\section{Conclusions}

\indent\indent Although the top-quark lifetime is much less than the
strong-interaction time scale, $\lqcd^{-1}$, there are nonperturbative
contributions to the top-quark pole mass, just as in the case of a
stable heavy quark.  These nonperturbative contributions are signaled
by the divergent behavior at large orders of an expansion in $a =
\beta_0\alpha_s(m_R)/4\pi$.  This leads to an unavoidable ambiguity of
$O(\lqcd)$ in the pole mass of the top quark.

A short-distance mass, such as the $\overline{\rm MS}$ mass, can in
principle be measured with arbitrary accuracy.  This may require
nonperturbative information, depending on the measurement.  It is
sensible to adopt the $\overline{\rm MS}$ mass as the standard
definition of the top-quark mass, as is the convention for the lighter
quarks \cite{PDG}.  The relation between the top-quark pole mass and
the $\overline{\rm MS}$ mass evaluated at the pole mass, $\overline
m(m_{pole})$, is known to two loops \cite{GBGS}:
\begin{equation}
m_{pole} = \overline m(m_{pole}) \left( 1 +\frac{4}{3}
\frac{\overline\alpha_s(m_{pole})}{\pi} + 10.95 \left(
\frac{\overline\alpha_s(m_{pole})}{\pi} \right)^2 + \cdots \right) +
O(\lqcd)
\end{equation}
where the last term reminds us that the pole mass has an unavoidable
ambiguity of $O(\lqcd)$.  Given that the pole mass is ambiguous, we
suggest as the standard the $\overline{\rm MS}$ mass evaluated at the
$\overline{\rm MS}$ mass, which is related to the pole mass by
\begin{equation}
m_{pole}=\overline m(\overline m) \left( 1 + \frac{4}{3}
\frac{\overline\alpha_s(\overline m)}{\pi} + 8.28 \left(
\frac{\overline\alpha_s(\overline m)}{\pi} \right)^2 + \cdots \right)
+ O(\lqcd) \;.
\end{equation}
The difference in the coefficients of the two ${\overline\alpha_s}^2$
terms above is exactly 8/3.  For a top-quark pole mass of $175 \pm 6$
GeV, $\overline m(\overline m) =166 \pm 6$ GeV.\footnote{$\overline
m(m_{pole})=165 \pm 6$ GeV.}

The considerations of this paper apply to any colored particle, stable
or unstable.  Thus, if nature is supersymmetric, the pole masses of
squarks and gluinos will necessarily be ambiguous by an amount
proportional to $\lqcd$.

\section*{Acknowledgements}

\indent\indent We are grateful for conversations with E.~Braaten,
A.~El-Khadra, R.~Leigh, T.~Liss, and T.~Stelzer. This work was
supported in part by Department of Energy grant DE-FG02-91ER40677. We
gratefully acknowledge the support of a GAANN fellowship, under grant
number DE-P200A10532, from the U.~S.~Department of Education for M.~S.


\begin{thebibliography}{99}

\bibitem{TOP} CDF Collaboration, F.~Abe~{\it et al.}, \PRL 74 2626 1995 ;
D0 Collaboration, S.~Abachi~{\it et al.}, \PRL 74 2632 1995 .

\bibitem{MASS} P.~Tipton, presented at the XXVIII International Conference
on High-Energy Physics, Warsaw, Poland, July 1996.

\bibitem{TEV2000} {\sl Future Electroweak Physics at the Fermilab Tevatron:
Report of the tev\_2000 Study Group}, eds.~D.~Amidei and R.~Brock, 
FERMILAB-Pub-96/082 (1996).

\bibitem{ATLAS} ATLAS Technical Proposal, CERN/LHCC/94-43, LHCC/P2 (1994).

\bibitem{NLC} {\sl Physics and Technology of the Next Linear Collider},
NLC ZDR Design Group and the NLC Physics Working Group,
hep-ex/9605011.

\bibitem{MUON} {\sl $\mu^+\mu^-$ Collider: A Feasibility Study},
$\mu^+\mu^-$ Collider Collaboration, BNL-52503 (1996).

\bibitem{T} R.~Tarrach, \NPB 183 384 1981 .

\bibitem{BSUV} I.~Bigi, M.~Shifman, N.~Uraltsev, and A.~Vainshtein, \PRD 
50 2234 1994 .

\bibitem{BB} M.~Beneke and V.~Braun, \NPB 426 301 1994 .

\bibitem{BU} I.~Bigi and N.~Uraltsev, \PLB 321 412 1994 .

\bibitem{K} J.~K\"uhn, Ann.~Phys.~Austr., Suppl.~XXIV, 203 (1982).

\bibitem{BDKKZ} I.~Bigi, Y.~Dokshitzer, V.~Khoze, J.~K\"uhn, and
P.~Zerwas, \PLB 181 157 1986 .

\bibitem{OR} L.~Orr and J.~Rosner, \PLB 246 221 1990 ; 
L.~Orr, \PRD 44 88 1991 .

\bibitem{FK} V.~Fadin and V.~Khoze, Pis'ma Zh. Eksp. Teoor. Fiz. {\bf
46}, 417 (1987) [JETP Lett. {\bf 46}, 525 (1987)]; Yad. Fiz. {\bf 48},
487 (1988) [Sov. J. Nucl. Phys., {\bf 48}, 309 (1988)]; V.~Fadin,
V.~Khoze, and T.~Sj\"{o}strand, \ZPC 48 613 1990 .

\bibitem{SP} M.~Strassler and M.~Peskin, \PRD 43 1500 1995 .

\bibitem{KS} V.~Khoze and T.~Sj\"{o}strand, \PLB 328 466 1994 .

\bibitem{ELOP1} R.~Eden, P.~Landshoff, D.~Olive, and J.~Polkinghorne,
{\sl The Analytic S-Matrix} (Cambridge University Press, Cambridge,
1966), section 4.5.

\bibitem{ELOP2} Ref.~\cite{ELOP1}, section 4.9.

\bibitem{NNA} M.~Beneke and V.~Braun, \PLB 348 513 1995 .

\bibitem{BBB} P.~Ball, M.~Beneke, and V.~Braun, \NPB 452 563 1995 .

\bibitem{PS} K.~Philippides and A.~Sirlin, \NPB 450 3 1995 . 

\bibitem{tH} G.~'t Hooft, in {\sl The Whys of Subnuclear Physics},
Proceedings of the International School of Subnuclear Physics, Erice,
1977, ed.~A.~Zichichi (Plenum, New York, 1979), p.~943; A.~Mueller, in
{\sl QCD - 20 Years Later}, Proceedings of the Workshop, Aachen,
Germany, 1992, eds.~P.~Zerwas and H.~Kastrup (World Scientific,
Singapore, 1993), Vol.~1, p.~162.

\bibitem{SD} Itzykson and Zuber, {\sl Quantum Field Theory}
(McGraw-Hill, New York, 1980), p.~475.

\bibitem{PDG} {\sl Review of Particle Properties}, \PRD 54 1 1996 , p.~303.

\bibitem{GBGS} N.~Gray, D.~Broadhurst, W.~Grafe, and K.~Schilcher,
\ZPC 48 673 1990 .

\end{thebibliography}
\end{document}